\pdfoutput=1
\documentclass[conference]{IEEEtran}
\IEEEoverridecommandlockouts
\usepackage{amsmath}
\usepackage{graphicx}
\usepackage{textcomp}
\usepackage{xcolor}
\usepackage[10pt]{moresize}
\usepackage{multirow}

\usepackage{amsthm}
\theoremstyle{definition}

\usepackage{algorithm}
\usepackage{algorithmicx}
\usepackage{algpseudocode}
\usepackage{comment}
\algdef{SE}[DOWHILE]{Do}{doWhile}{\algorithmicdo}[1]{\algorithmicwhile\ #1}%

\usepackage{epsfig}

\usepackage[font=footnotesize,skip=6pt]{caption}
\usepackage{subcaption}

\begin{document}
% \IEEEoverridecommandlockouts
% \IEEEpubid{\makebox[\columnwidth]{ 979-8-3503-7608-1/24\$31.00 \copyright2024 IEEE \hfill} \hspace{\columnsep}\makebox[\columnwidth]{ }}

\title{Large Language Model (LLM) for Standard Cell Layout Design Optimization}

\vspace{-5cm}
\author{\IEEEauthorblockN{Chia-Tung Ho}
\IEEEauthorblockA{NVIDIA Research \\ Santa Clara, CA, USA\\
chiatungh@nvidia.com}
\and
\IEEEauthorblockN{Haoxing Ren}
\IEEEauthorblockA{NVIDIA Research \\ Austin, TX, USA\\
haoxingr@nvidia.com}}

\maketitle
\vspace{-5cm}

\begin{abstract}
Standard cells are essential components of modern digital circuit designs. With process technologies advancing toward 2{\em nm}, more routability issues have arisen due to the decreasing number of routing tracks, increasing number and complexity of design rules, and strict patterning rules. 
The state-of-the-art standard cell design automation framework is able to automatically design standard cell layouts in advanced nodes, but it is still struggling to generate highly competitive Performance-Power-Area (PPA) and routable cell layouts for complex sequential cell designs.
Consequently, a novel and efficient methodology incorporating the expertise of experienced human designers to incrementally optimize the PPA of cell layouts is highly necessary and essential.
% As a result, a novel and efficient methodology with experienced human designers' expertise to optimize the PPA of cell layouts incrementally is highly needed and essential.

\textcolor{black}{High-quality device clustering, with consideration of netlist topology, diffusion sharing/break and routability in the layouts, can reduce complexity and assist in finding highly competitive PPA, and routable layouts faster.
In this paper, we leverage the natural language and reasoning ability of Large Language Model (LLM) to generate high-quality cluster constraints incrementally to optimize the cell layout PPA and debug the routability with ReAct prompting.
\color{black}{On a benchmark of sequential standard cells in 2{\em nm}, we demonstrate that the proposed method not only achieves up to 19.4\% smaller cell area, but also generates 23.5\% more LVS/DRC clean cell layouts than previous work.
In summary, the proposed method not only successfully reduces cell area by 4.65\% on average, but also is able to fix routability in the cell layout designs.}}

\end{abstract}
\vspace{-0.25cm}
% \begin{IEEEkeywords}
% standard cell, Complementary-FET (CFET), automated cell generation, placement, routing, cell synthesis, SMT, design technology co-optimization (DTCO), system technology co-optimization (STCO).
% \end{IEEEkeywords}

\IEEEpeerreviewmaketitle

\section{Introduction}
Standard cells are essential components of modern digital circuit designs. As process technologies relentlessly advance toward 2{\em nm}, designing a cell with competitive Performance-Power-Area (PPA) while considering routability becomes increasingly challenging due to the decreasing number of routing tracks, increasing complexity of design rules, and strict patterning rules.
The state-of-the-art standard cell design automation framework is capable of automatically designing standard cell layouts in advanced nodes, but it still struggles to generate highly competitive PPA and routable cell layouts for complex sequential cell designs.
As a result, a novel and efficient methodology, leveraging the expertise of experienced human designers, to optimize the PPA and routability of cell layouts incrementally, has emerged as a critical need.

Recently, automated standard cell synthesis tools such as NVCell~\cite{ren2021nvcell} and BonnCell~\cite{van2019bonncell}, have been shown to generate high quality cell layouts on advanced technology nodes. Due to routability issues, one of the key challenges is that the generated placement for any given cell could be unroutable or unable to be routed without DRC errors. 
%Previous works use simpler heuristics or models to predict routability, but they are still struggling to generate routable device placements when considering complex design rules and multi-patterning rules in the actual routing stage. 
NVCell2~\cite{ho2023nvcell} develops a lattice graph routability model and successfully improves the routability in the advanced technology nodes. However, its performance is not scale to hundreds of transistors because the model inference needs to be performed for every action in the simulated annealing-based placement algorithm~\cite{ren2021nvcell} and the cell-level metrics (i.e., cell width (CW) and total wirelength (TWL)) are compromised for routability. 
Ho {\em et al.}~\cite{ho2024novel} proposed a transformer model-based cluster approach to generate high-quality device cluster constraints, which considers diffusion sharing/break, routability, and DRCs of routing metals in the layout of different technology nodes, and achieved better PPA, routability, and performance than~\cite{ho2023nvcell} in the advanced nodes.
However, selecting a good set of LVS/DRC clean layouts to train the transformer cluster model for optimizing PPA and routability of complex sequential cells together is quite challenging. This is because cells with routability issues typically have a larger cell width to reduce transistor pin density, while cells with a more compact layout could exacerbate routability issues. Additionally, there is a limited amount of LVS/DRC clean layouts available for training the transformer cluster model in the early development stage of cell library in a new technology node.

\begin{figure} [!t]
	\centering
	\includegraphics[width=0.9\columnwidth]{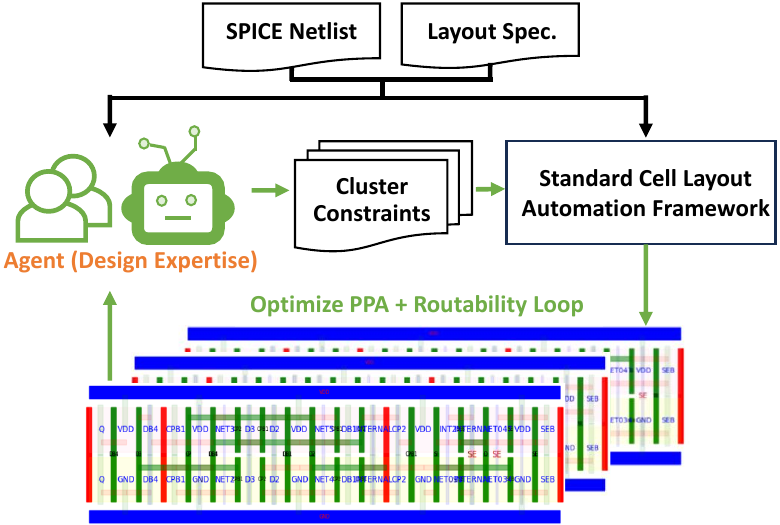}
    \vspace{-0.2cm}
	\caption{An illustration of the PPA and routability optimization loop includes an agent with design expertise, adjusted cluster constraints, and a standard cell layout automation framework (i.e., NVCell).}
	\label{AgentOptLoopFig}
\vspace{-0.5cm}
\end{figure}

An agent with designers' expertise can adjust and fine-tune device clustering constraints incrementally based on the netlist and the layout of the previous iteration to efficiently optimize PPA and routability together as shown in \textcolor{black}{Figure~\ref{AgentOptLoopFig}}. 
Lately, Large Language Models (LLMs) have shown great promise across various tasks in language understanding and interactive decision-making, incorporating reasoning and actions. 
In this paper, we leverage the natural language and reasoning ability of LLMs to adjust the device clustering constraints incrementally, optimizing cell layout PPA and routability with guidance from designers' expertise and ReAct~\cite{yao2022react} prompting techniques. Our main contributions are as follows.
\begin{itemize}
    \item  \textcolor{black}{We are the first to explore LLM for optimization in Electronic Design Automation (EDA) on an industrial-level benchmark.} We propose a novel, and efficient LLM for standard cell layout design optimization methodology to generate high-quality cluster constraints to optimize the cell layout PPA and debug the routability with guidance of designers' expertise and ReAct~\cite{yao2022react} prompting techniques on an industrial technology node.
    The proposed methodology can improve the cell-level PPA and generate the device clusters incrementally considering netlist, previous cluster constraints, routability, and physical layout, simultaneously.
    
    \item We conduct holistic assessments and studies on the capabilities and domain knowledge of existing LLM on SPICE netlist language, cluster design constraint format, and physical layout description. Then, we automate the domain knowledge extraction with guidelines from designers' expertise for optimizing the PPA and routability together.

    \item  The proposed novel LLM for standard cell layout design optimization methodology achieves up to 19.4\% smaller cell area, and generates 23.5\% more LVS/DRC clean cell layouts than a state-of-the-art baseline on a benchmark of sequential standard cells in industrial 2{\em nm} technology node.
\end{itemize}
The remaining sections are organized as follows: Section~\ref{DomainKnowledgeSection} demonstrates the study and assessment of existing LLM on understanding the netlist and standard cell layout domain knowledge. Section~\ref{MethodologySection} describes our novel LLM for standard cell layout design optimization methodology. Section~\ref{ExperimentalSection} presents our main experiment. Section~\ref{ConclusionSection} concludes the paper.

\vspace{-0.1cm}
\section{Standard Cell Layout Design Domain Knowledge} \label{DomainKnowledgeSection}
% \vspace{-0.2cm}
We conduct assessments and studies on the capabilities and domain knowledge of existing LLMs of SPICE language format, device cluster constraints, and physical layout of standard cells. 
% To utilize LLMs for standard cell layout design optimization, it is essential to consider the connections of the devices in the netlist, the design cluster constraint format, and coordinate-based physical layout information.

\begin{figure} [!t]
	\centering
	\includegraphics[width=0.88\columnwidth]{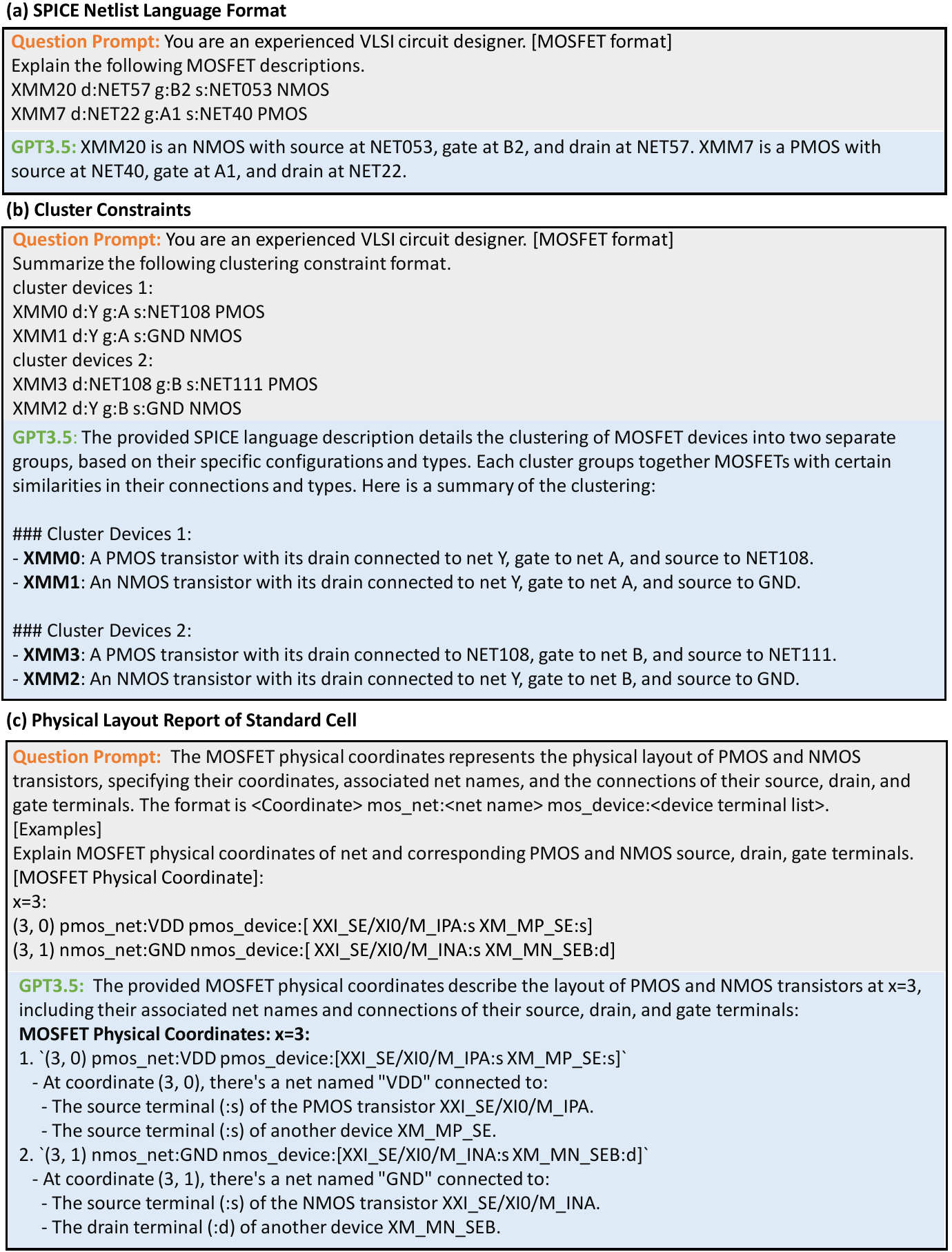}
    \vspace{-0.1cm}
	\caption{Standard cell layout design domain knowledge assessment of existing LLM on: (a) SPICE language, (b) cluster constraint, and (c) physical layout.}
	\label{DomainKnowledgeFig}
    \vspace{-0.2cm}
\end{figure}

\vspace{-0.1cm}
\subsection{SPICE Netlist Language}
To assess the domain knowledge of LLM on SPICE netlist language, we input the technology independent device (i.e., MOSFET) description in SPICE language format and let LLM explain the MOSFET information. The technology independent description of a MOSFET includes name, terminal connections, and the type of MOSFET. 
% The format is described as "NAME s:SOURCE g:GATE d:DRAIN TYPE". 
Figure~\ref{DomainKnowledgeFig}(a) shows that LLM can understand technology independent MOSFET description in SPICE language format and identify the net connection information at the drain, gate, and source terminals\footnote{The [MOSFET format] is "A MOSFET can be described in SPICE format as MOSFET\_NAME d:DRAIN g:GATE s:SOURCE MOSFET\_TYPE".}. 
%As a result, the technology independent device description in SPICE language format can be leveraged to describe the standard cell netlist topology information.

\vspace{-0.2cm}
\subsection{Cluster Constraint}
We study the ability of existing LLM on understanding the cluster design constraint format including multiple devices and clusters. We show one of the example studies in Figure~\ref{DomainKnowledgeFig}(b). 
We task the LLM with summarizing the provided clusters and the devices associated with them for subsequent reasoning tasks. The study's findings indicate that the existing LLM can accurately identify the total number of clusters and the information regarding devices within these clusters.

\vspace{-0.2cm}
\subsection{Standard Cell Layout}
The standard cell layout includes the placed device locations and the net connection of device terminals. We use coordinates and the corresponding device and its terminals to represent the standard cell layout, as shown in Figure~\ref{DomainKnowledgeFig}(c). We study the capability of LLM in understanding the location of placed devices and transistor terminal connections for the standard cell layout design optimization task. In one of the study examples in Figure~\ref{DomainKnowledgeFig}(c), the LLM successfully explains the placed MOSFETs and their terminals at each coordinate.

In summary, the existing LLM (i.e., GPT3.5) has the capability to understand the netlist topology through SPICE language, device cluster constraint, and device placement, and the terminal connections in standard cell layout description for generating better device cluster constraints for PPA and routability debugging.

\begin{figure} [!t]
	\centering
	\includegraphics[width=0.95\columnwidth]{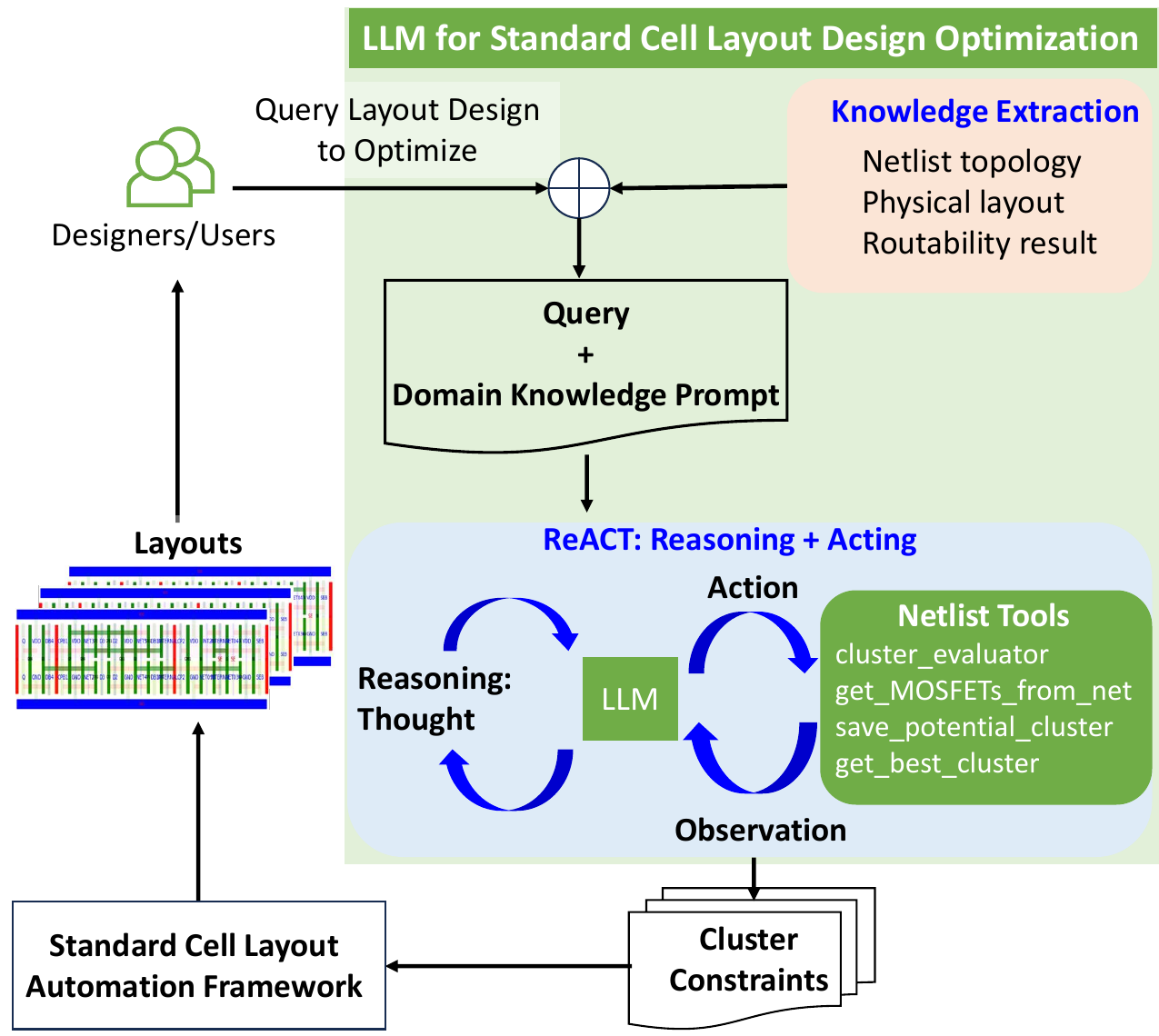}
	\caption{Overview of LLM for Standard cell layout design optimization flow}
	\label{LLMReAct_frameworkFig}
    \vspace{-0.3cm}
\end{figure}

\vspace{-0.2cm}
\section{LLM for Standard Cell Layout Design Optimization} \label{MethodologySection}
We introduce the details of LLM for standard cell layout design optimization, which leverages the natural language and reasoning ability of LLM to generate high-quality cluster constraints to optimize the cell-level PPA and debug routability issues.
The overall flow is outlined in Section~\ref{MethodologySection:Overview}, and Figure~\ref{LLMReAct_frameworkFig}.
Section~\ref{MethodologySection:Tools} introduces the developed netlist tools, which provide accurate sub-circuit retrieval from the netlist and cluster evaluation, for assisting LLM on layout design optimization task. 
Finally, we discuss the application of ReAct~\cite{yao2022react} in Section~\ref{MethodologySection:ReAct}.

\vspace{-0.2cm}
\subsection{Flow Overview} \label{MethodologySection:Overview}
\vspace{-0.1cm}
The proposed LLM for standard cell layout design optimization comprises the following components: knowledge extraction to initiate queries and provide domain knowledge prompts, netlist tools to assist the LLM in generating valid cluster constraints, and reasoning and action in ReAct~\cite{yao2022react} for exploring high-quality cluster candidates for the PPA/routability of layout design.

Fristly, designers input the initial layout and its corresponding cluster constraints. 
The knowledge extraction is used to create the domain knowledge prompt with netlist topology in technology independent descriptions of MOSFETs, initial cluster constraints, standard cell layout, and guidance from designers' expertise as described in Section~\ref{DomainKnowledgeSection}. 
Then, the ReAct prompting method allows LLM to perform dynamic reasoning to create and adjust plans for acting (reason to act), while also interacting with the netlist tools (i.e., grouping MOSFET, evaluating clusters, etc.) to incorporate additional information into reasoning (act to reason).
These cluster constraints can be fed into NVCell~\cite{ho2023nvcell} to generate layouts. 
Designers can repeat this process until the PPA meets the requirements without rouability issues.  

\vspace{-0.2cm}
\subsection{Netlist Tools} \label{MethodologySection:Tools}
\vspace{-0.1cm}
We have developed a set of netlist tools to assist LLM in generating valid cluster constraints and correctly retrieving sub-circuits in the ReAct reasoning and action loop.
The netlist tools include a cluster evaluator, a function to retrieve group devices from nets, a mechanism to save potential clusters, and a function to obtain the best cluster result.

\noindent {\bf Cluster evaluator}:
This tool is used to evaluate the quality of the generated cluster result using simple cluster score to capture the potential diffusion sharing and common gate in the layout. 
Here, we use the simple cluster score for evaluation in ReAct because the turnaround time is too long to launch layout generation to collect accurate cell layout metrics (i.e., CW, TWL, etc.).
The simple cluster score is calculated in Equation~(\ref{SimpleMetricEq}). 
A larger score means the devices can potentially be placed with more common diffusion sharing and common gate inside each cluster.

\vspace{-0.3cm}
\begin{equation} \label{SimpleMetricEq}
\scriptsize{
\begin{aligned}
cluster\_score = \sum_{c \in {\bf C}}(\frac{\sum_{n \in \bf{N^d_c}} \lfloor\frac{P_n}{2} \rfloor + \lfloor\frac{N_n}{2}\rfloor}{T_c} + \frac{\sum_{n \in \bf{N^g_c}} \min(P_n, N_n)}{T_c})
\end{aligned}
}
\end{equation}
Where $c$ is a cluster in a set of clusters, $\bf{C}$. $\bf{N^d_c}$ and $\bf{N^g_c}$ represent the set of nets at diffusion (i.e., source or drain terminals) and gate in cluster $c$. $P_n$ and $N_n$ represent the number of net $n$ at PMOS and NMOS terminals, respectively. $T_c$ is the number of transistors in cluster $c$. For potential diffusion sharing in a cluster, we calculate the number of same net pairs at diffusion terminals for PMOS and NMOS, respectively. The minimum number of the same gate net of PMOS and NMOS is the potential number of common gates in a cluster.

\noindent {\bf Get group devices from nets}: This netlist tool returns the group of transistors from an arbitrary number of nets in the netlist. LLM can use this tool to search and explore potential device clusters.

\noindent {\bf Save potential cluster}:
This tool returns the current clusters and cluster score after LLM inputs a new potential cluster. The duplicated devices in different clusters are fixed based on the number of shared nets of these duplicated devices in each cluster.

\noindent {\bf Get best cluster result}:
This tool returns the cluster result with the best simple cluster score (i.e., Equation~(\ref{SimpleMetricEq})). It helps LLM revert back or restart the search from the previous best cluster result when it is stuck in the searching potential cluster phase.

\begin{figure} [!t]
	\centering
	\includegraphics[width=0.95\columnwidth]{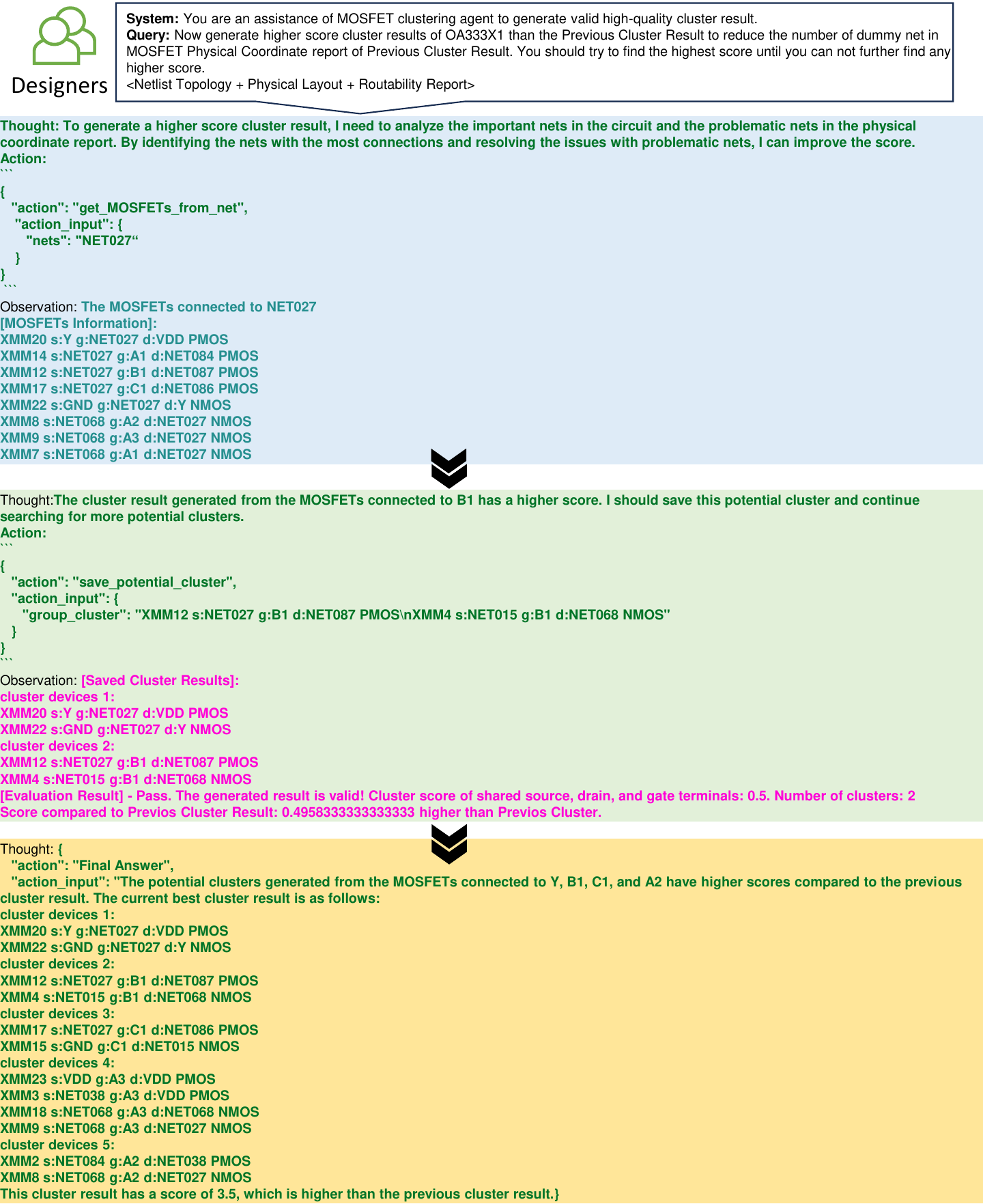}
        \vspace{-0.2cm}
	\caption{An example of ReAct steps of Thought-Action-Observation for standard cell layout design optimization.}
	\label{ReActExampleFig}
\vspace{-0.3cm}
\end{figure}

\vspace{-0.2cm}
\subsection{ReAct: Reason + Act Loop}\label{MethodologySection:ReAct}
\vspace{-0.1cm}
We enable LLMs to function as autonomous circuit design agent for reasoning and acting with the netlist tools through {\bf{ReAct}} prompting mechanism~\cite{yao2022react}. In ReAct, the LLM initiates the generation of subsequent steps with Thought, Action, and Observation components. 
The action is querying one of the netlist tools described in Section~\ref{MethodologySection:Tools}. The output of the queried netlist tool from action becomes the observation in the prompt.
The agent continues the reasoning and acting traces until selecting the "Final Answer" action.
Figure~\ref{ReActExampleFig} shows an example of optimizing the cell area of a standard cell. 
\textcolor{black}{Here, the agent starts with querying the group of devices connected to NET027 to explore good clusters incrementally to reduce the diffusion break for area reduction since NET027 is one of the high connection nets in the netlist topology and abutted to the diffusion break dummy device in the physical layout.
Finally, the agent successfully generates high-quality cluster result through reasoning and leveraging the netlist tools traces in ReAct.}

\begin{figure} [!b]
	\centering
	\includegraphics[width=1\columnwidth]{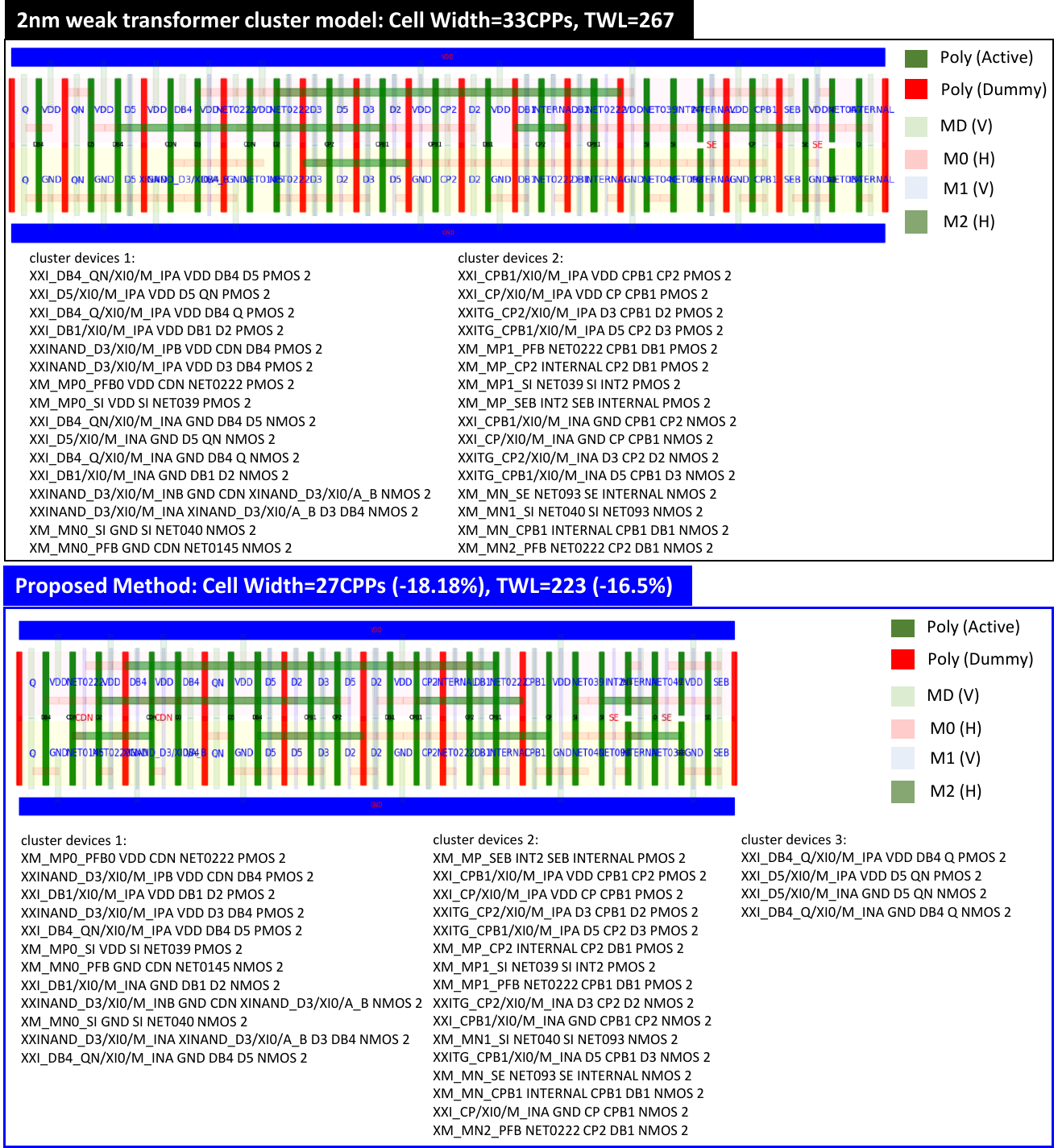}
 \vspace{-0.6cm}
	\caption{The cluster constraints and cell layouts of {\bf 2{\em nm} weak transformer cluster model}~\cite{ho2024novel}, and proposed method of Seq3 cell. The proposed method successfully reduce the CW, and TWL by 18.18\%, and 16.48\%, respectively.}
	\label{SDFCND1LayoutFig}
\vspace{-0.2cm}
\end{figure}
% \usepackage{multirow}
\begin{table}[!b]
\caption{CW (CPPs) table of the 17 complex sequential cells in 2{\em nm} of transformer cluster model~\cite{ho2024novel}, simulated annealing (SA), and the proposed method. The SA algorithm is implemented base on~\cite{cicirello2007design}.
\#Devs=number of transistors.
Opt.(a)/(b)=Optimize layouts from (a)/(b).
X=unroutable. V=Fix routability. Success(\%)=the percentage of LVS/DRC clean cell layouts.} \label{CellWidthTable}
\vspace{-0.1cm}
\tabcolsep = 1.2pt
\centering
\scriptsize{
\begin{tabular}{|cr|rr|rr|rrrr|}
\hline
\multicolumn{1}{|c|}{\multirow{3}{*}{Cell}} & \multicolumn{1}{c|}{\multirow{3}{*}{\#Devs}} & \multicolumn{2}{c|}{Transformer Cluster~\cite{ho2024novel}}                                                           & \multicolumn{2}{c|}{SA}                                                                         & \multicolumn{4}{c|}{Proposed Method}                                                                                                                            \\ \cline{3-10} 
\multicolumn{1}{|c|}{}                      & \multicolumn{1}{c|}{}                           & \multicolumn{1}{c|}{\multirow{2}{*}{(a) 2nm weak}} & \multicolumn{1}{c|}{\multirow{2}{*}{(b) 5nm}} & \multicolumn{1}{c|}{\multirow{2}{*}{Opt. (a)}} & \multicolumn{1}{c|}{\multirow{2}{*}{Opt. (b)}} & \multicolumn{1}{c|}{\multirow{2}{*}{Opt. (a)}} & \multicolumn{1}{c|}{\multirow{2}{*}{Opt. (b)}} & \multicolumn{2}{c|}{Impr. (\%)}                               \\ \cline{9-10} 
\multicolumn{1}{|c|}{}                      & \multicolumn{1}{c|}{}                           & \multicolumn{1}{c|}{}                              & \multicolumn{1}{c|}{}                         & \multicolumn{1}{c|}{}                          & \multicolumn{1}{c|}{}                          & \multicolumn{1}{c|}{}                          & \multicolumn{1}{c|}{}                          & \multicolumn{1}{c|}{Over (a)} & \multicolumn{1}{c|}{Over (b)} \\ \hline
\multicolumn{1}{|c|}{Seq1}                  & 40                                              & \multicolumn{1}{r|}{26}                            & 31                                            & \multicolumn{1}{r|}{X}                         & 31                                             & \multicolumn{1}{r|}{25}                        & \multicolumn{1}{r|}{25}                        & \multicolumn{1}{r|}{3.85}     & 19.35                         \\ \hline
\multicolumn{1}{|c|}{Seq2}                  & 60                                              & \multicolumn{1}{r|}{X}                             & 41                                            & \multicolumn{1}{r|}{X}                         & X                                              & \multicolumn{1}{r|}{39}                        & \multicolumn{1}{r|}{39}                        & \multicolumn{1}{r|}{V}        & 4.88                          \\ \hline
\multicolumn{1}{|c|}{Seq3}                  & 40                                              & \multicolumn{1}{r|}{33}                            & 26                                            & \multicolumn{1}{r|}{27}                        & 26                                             & \multicolumn{1}{r|}{27}                        & \multicolumn{1}{r|}{26}                        & \multicolumn{1}{r|}{18.18}    & 0.00                          \\ \hline
\multicolumn{1}{|c|}{Seq4}                  & 38                                              & \multicolumn{1}{r|}{27}                            & 25                                            & \multicolumn{1}{r|}{X}                         & 25                                             & \multicolumn{1}{r|}{25}                        & \multicolumn{1}{r|}{25}                        & \multicolumn{1}{r|}{7.41}     & 0.00                          \\ \hline
\multicolumn{1}{|c|}{Seq5}                  & 36                                              & \multicolumn{1}{r|}{23}                            & 22                                            & \multicolumn{1}{r|}{26}                        & 22                                             & \multicolumn{1}{r|}{22}                        & \multicolumn{1}{r|}{22}                        & \multicolumn{1}{r|}{4.35}     & 0.00                          \\ \hline
\multicolumn{1}{|c|}{Seq6}                  & 36                                              & \multicolumn{1}{r|}{22}                            & 25                                            & \multicolumn{1}{r|}{22}                        & 22                                             & \multicolumn{1}{r|}{22}                        & \multicolumn{1}{r|}{22}                        & \multicolumn{1}{r|}{0.00}     & 12.00                         \\ \hline
\multicolumn{1}{|c|}{Seq7}                  & 34                                              & \multicolumn{1}{r|}{20}                            & 20                                            & \multicolumn{1}{r|}{20}                        & 20                                             & \multicolumn{1}{r|}{20}                        & \multicolumn{1}{r|}{20}                        & \multicolumn{1}{r|}{0.00}     & 0.00                          \\ \hline
\multicolumn{1}{|c|}{Seq8}                  & 38                                              & \multicolumn{1}{r|}{25}                            & 25                                            & \multicolumn{1}{r|}{25}                        & 25                                             & \multicolumn{1}{r|}{24}                        & \multicolumn{1}{r|}{25}                        & \multicolumn{1}{r|}{4.00}     & 0.00                          \\ \hline
\multicolumn{1}{|c|}{Seq9}                  & 32                                              & \multicolumn{1}{r|}{19}                            & 19                                            & \multicolumn{1}{r|}{19}                        & 19                                             & \multicolumn{1}{r|}{19}                        & \multicolumn{1}{r|}{19}                        & \multicolumn{1}{r|}{0.00}     & 0.00                          \\ \hline
\multicolumn{1}{|c|}{Seq10}                 & 34                                              & \multicolumn{1}{r|}{20}                            & 21                                            & \multicolumn{1}{r|}{20}                        & 20                                             & \multicolumn{1}{r|}{20}                        & \multicolumn{1}{r|}{20}                        & \multicolumn{1}{r|}{0.00}     & 4.76                          \\ \hline
\multicolumn{1}{|c|}{Seq11}                 & 40                                              & \multicolumn{1}{r|}{26}                            & 26                                            & \multicolumn{1}{r|}{25}                        & 28                                             & \multicolumn{1}{r|}{25}                        & \multicolumn{1}{r|}{25}                        & \multicolumn{1}{r|}{3.85}     & 3.85                          \\ \hline
\multicolumn{1}{|c|}{Seq12}                 & 38                                              & \multicolumn{1}{r|}{24}                            & 27                                            & \multicolumn{1}{r|}{26}                        & 28                                             & \multicolumn{1}{r|}{24}                        & \multicolumn{1}{r|}{24}                        & \multicolumn{1}{r|}{0.00}     & 11.11                         \\ \hline
\multicolumn{1}{|c|}{Seq13}                 & 56                                              & \multicolumn{1}{r|}{X}                             & X                                             & \multicolumn{1}{r|}{X}                         & X                                              & \multicolumn{1}{r|}{41}                        & \multicolumn{1}{r|}{40}                        & \multicolumn{1}{r|}{V}        & V                             \\ \hline
\multicolumn{1}{|c|}{Seq14}                 & 44                                              & \multicolumn{1}{r|}{X}                             & 36                                            & \multicolumn{1}{r|}{X}                         & X                                              & \multicolumn{1}{r|}{34}                        & \multicolumn{1}{r|}{34}                        & \multicolumn{1}{r|}{V}        & 5.56                          \\ \hline
\multicolumn{1}{|c|}{Seq15}                 & 42                                              & \multicolumn{1}{r|}{X}                             & 32                                            & \multicolumn{1}{r|}{X}                         & X                                              & \multicolumn{1}{r|}{31}                        & \multicolumn{1}{r|}{31}                        & \multicolumn{1}{r|}{V}        & 3.13                          \\ \hline
\multicolumn{1}{|c|}{Seq16}                 & 56                                              & \multicolumn{1}{r|}{42}                            & 40                                            & \multicolumn{1}{r|}{X}                         & 40                                             & \multicolumn{1}{r|}{35}                        & \multicolumn{1}{r|}{35}                        & \multicolumn{1}{r|}{16.67}    & 12.50                         \\ \hline
\multicolumn{1}{|c|}{Seq17}                 & 42                                              & \multicolumn{1}{r|}{25}                            & 25                                            & \multicolumn{1}{r|}{25}                        & 25                                             & \multicolumn{1}{r|}{25}                        & \multicolumn{1}{r|}{25}                        & \multicolumn{1}{r|}{0.00}     & 0.00                          \\ \hline
\multicolumn{2}{|c|}{Success (\%)}                                                       & \multicolumn{1}{r|}{76.50}                         & 94.10                                         & \multicolumn{1}{r|}{58.80}                     & 76.50                                          & \multicolumn{1}{r|}{100}                       & \multicolumn{1}{r|}{100}                       & \multicolumn{1}{r|}{-}        & -                             \\ \hline
\end{tabular}
}
\end{table}

\vspace{-0.2cm}
\section{Experimental Results} \label{ExperimentalSection}
\vspace{-0.1cm}
Our work is implemented with Python and LangChain~\cite{Chase_LangChain_2022}.
We conduct all experiments with gpt-3.5-turbo-16k-0613 as the LLM through OpenAI APIs~\cite{openai}. We set the sampling temperature of LLM to 0.1. For ReAct prompting, we restrict the LLM to a maximum of 15 iterations of Thought-Action-Observation. 

We conduct extensive studies on the cell area and routability using 17 complex sequential standard cells in industrial 2{\em nm} technology node.
% to demonstrate the cell area optimization and routability fixing capabilities of our proposed method.
Due to a lack of cell layout data in the early stage development of cell library in 2{\em nm}, we trained (a) {\bf 2{\em nm} weak transformer cluster model}~\cite{ho2024novel} using 124 LVS/DRC clean cell layouts, and (b) {\bf 5{\em nm} transformer cluster model}~\cite{ho2024novel} using 512 LVS/DRC clean cell layouts as the baselines. 
\textcolor{black}{
We implemented a simulated annealing algorithm (SA), which is based on the modified Lam annealing scheduler~\cite{cicirello2007design} that required no hyper parameter tuning, for comparing the efficiency of the proposed method. In SA, we sample arbitrary 1 to $k$ nets considering the weights of nets and querying a group of devices from {\bf Get group devices from nets} tool for each {\em action}. The unrouted nets, and nets abutted to diffusion break dummy device have higher weights. Then, this queried group of devices will be accepted or rejected to be saved using {\bf Save potential cluster} tool based on the temperature and the delta simple cluster score. 
Finally, we apply the proposed method, and SA to optimize the cell area and routability from the initial cluster constraints and cell layouts of (a), and (b) on the selected unseen 17 complex sequential cells. For the proposed method, we launch 10 optimization runs for each cell and select valid cluster results to generate cell layout using NVCell~\cite{ho2023nvcell}.}

Table~\ref{CellWidthTable} shows the CW (CPPs) of the selected 17 sequential cells of transformer cluster method~\cite{ho2024novel}, SA, and the proposed method. 
\textcolor{black}{SA fails to optimize the cell area and fix the routability because the simple cluster score can not fully capture diffusion sharing, and routability in the layout. It requires more holistic and efficient methodology for cluster exploration than a naive net sampling-based approach.}  
Compared to (a), the proposed method can not only reduce up to 18.18\% of CW, but also improve the success rate from 76.5\% to 100\%. In addition, the proposed method reduces up to 19.35\% of CW, and improves the success rate from 94.10\% to 100\% when optimizing the initial cell layout from (b). \textcolor{black}{Figure~\ref{SDFCND1LayoutFig} shows the layouts and generated cluster constraints of (a), and using proposed method to optimize from (a) of Seq3 cell. The proposed method achieves 18.18\% reduction in CW, and 16.48\% reduction in TWL.}

In summary, excluding the fix routability cells, the proposed method achieves 4.48\%, and 4.82\% reduction in cell area on average over (a), and (b). 
\textcolor{black}{We successfully demonstrate that the proposed method incorporates holistic information from the netlist and layout of complex sequential cells to conduct efficient cluster exploration with a simple cluster score for the standard cell layout design optimization task.}

\vspace{-0.2cm}
\section{Conclusion} \label{ConclusionSection}
\vspace{-0.1cm}
We propose a novel, efficient, and the first LLM for standard cell layout design optimization methodology to generate high-quality cluster constraints to optimize the cell layout PPA and debug the routability with the guidance of designers’ expertise and ReAct~\cite{yao2022react} prompting techniques.
We have demonstrated that the proposed method achieves up to 19.4\% smaller cell area, and generates 100\% LVS/DRC clean cell layouts on the selected complex sequential cell benchmark in 2{\em nm}.
This research not only provides a novel autonomous LLM agent for standard cell layout design optimization and debugging but also \textcolor{black}{introduces the new application of using LLM assistance for optimization in the EDA field for further exploration}.

\bibliographystyle{unsrt}
%{\tiny
\bibliography{main}
\newpage
\section{Appendix}
We present examples of netlist topology, physical layout, and routability report prompts from proposed knowledge extraction component as mentioned in Fig.~\ref{ReActExampleFig}. We describe the netlist topology and physical layout prompts of the simple cell example (i.e., OA333X1) in Fig.~\ref{ReActExampleFig}. For the routability report prompt, we select the Seq13 in Table~\ref{CellWidthTable}
for demonstration, as there are no unrouted nets in the OA333X1 layout depicted in Fig.~\ref{ReActExampleFig}.

\vspace{-0.3cm}
\begin{figure} [!b]
	\centering
	\includegraphics[width=0.8\columnwidth]{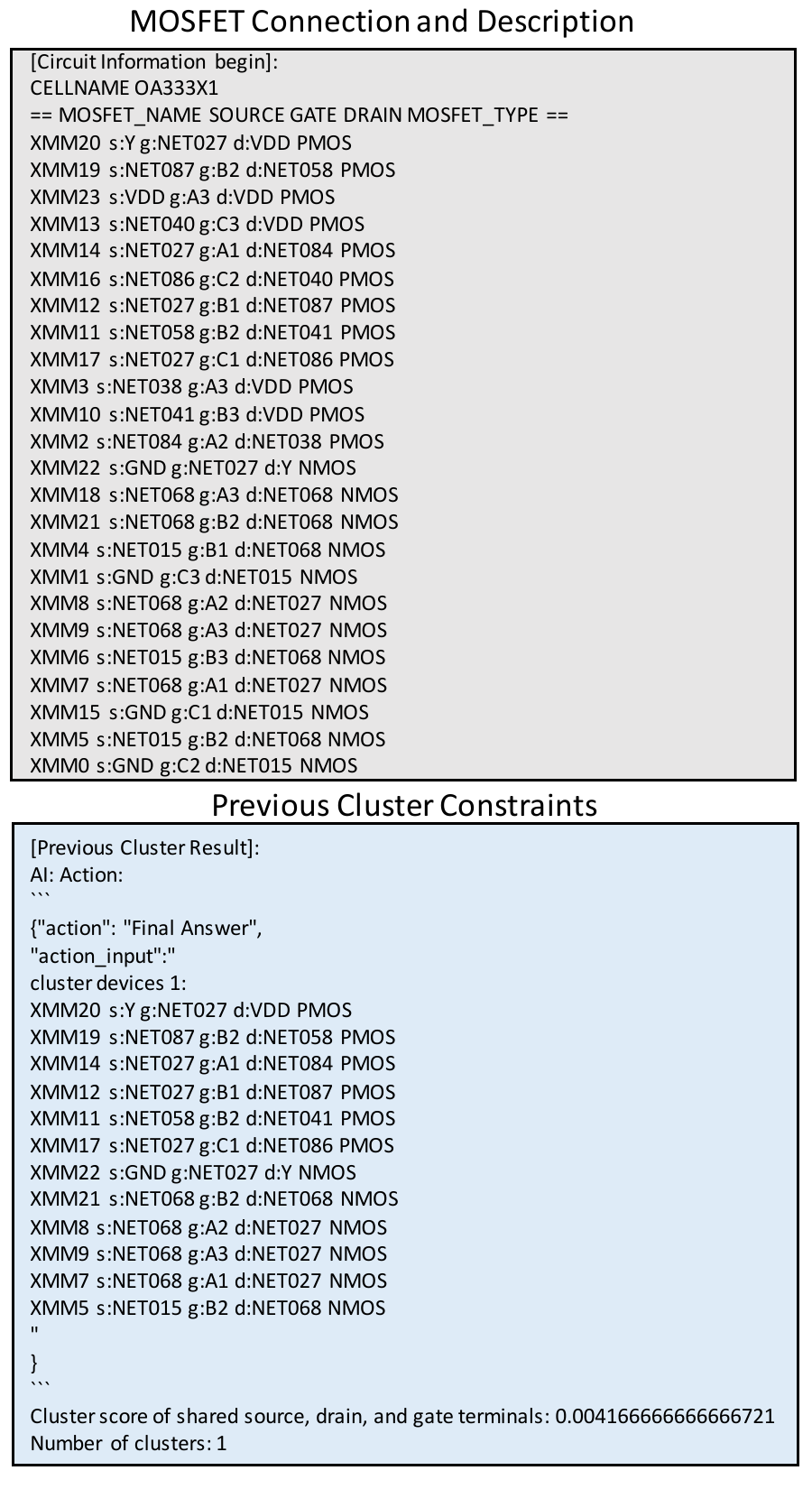}
	\caption{An example of extracted netlist topology prompt of the simple cell example in Fig.~\ref{ReActExampleFig}.}
	\label{NetlistPromptFig}
    \vspace{-0.3cm}
\end{figure}

% \vspace{-0.1cm}
\subsection{Netlist topology prompt}
The netlist topology prompt consists of MOSFET connection and description, as well as previous cluster constraints, as shown in Fig.~\ref{NetlistPromptFig}. In the MOSFET connection and description, each MOSFET is described using the technology-independent device description format introduced in Section~\ref{DomainKnowledgeSection}. 
For the previous cluster constraints, we present them in JSON BLOB format with the action labeled as "Final Answer".
At the end, we provide the simple cluster score and the number of clusters resulting from the previous cluster constraints.

With the netlist topology prompt information, the LLM agent can understand the netlist connections of each device, previous cluster constraints, and the simple cluster score of previous cluster constraints for the following reasoning and action for cell layout design optimization.

\subsection{Physical layout prompt}
We design the physical layout prompt with the placed device locations and net connection of device terminals for LLM to compile the netlist topology and layout together for ReAct steps of Thought-Action-Observation.

\begin{figure} [!h]
	\centering
	\includegraphics[width=1\columnwidth]{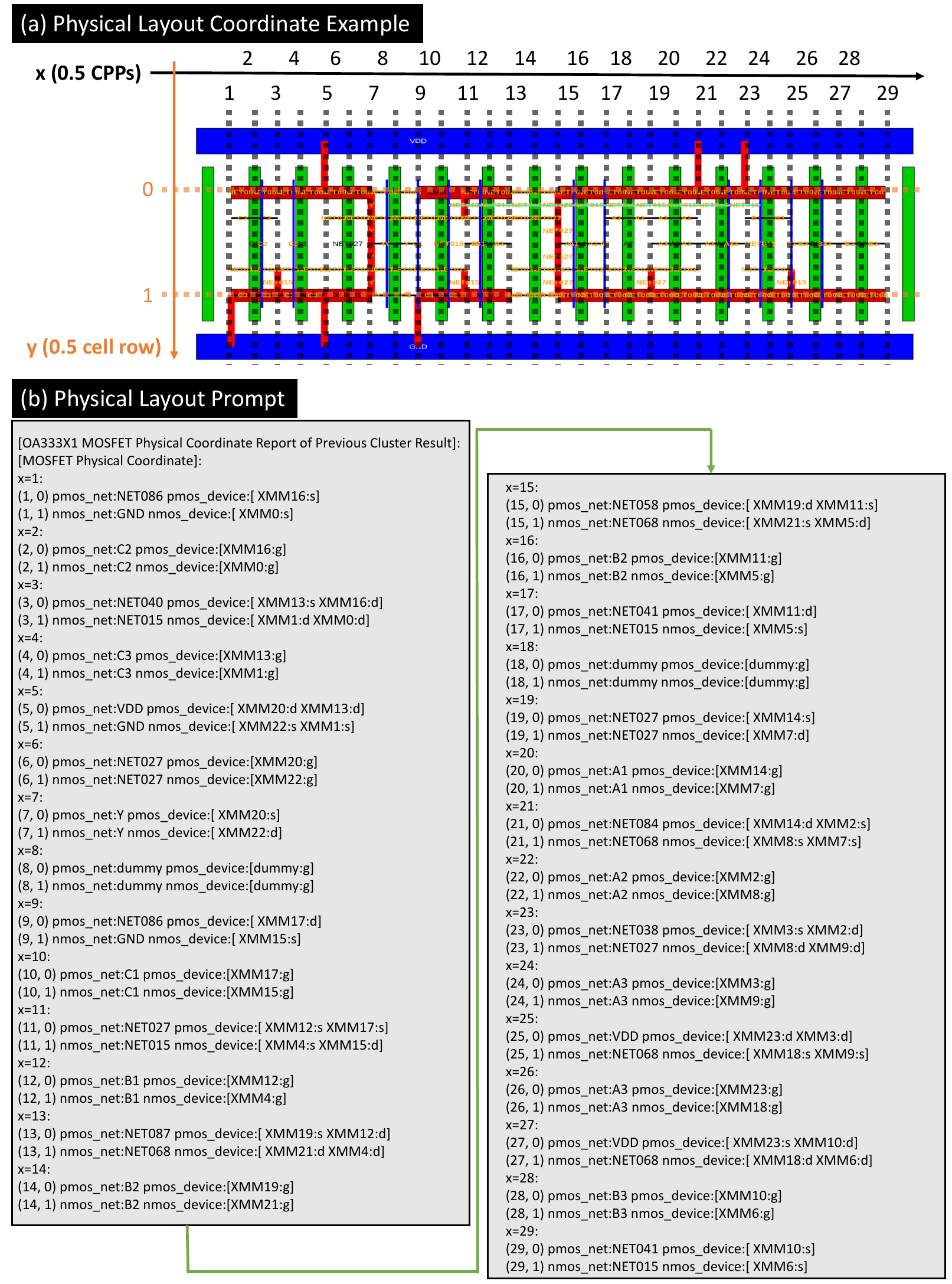}
	\caption{An example of extracted physical layout prompt of the simple cell example in Fig.~\ref{ReActExampleFig}.}
	\label{PhysicalLayoutPromptFig}
    \vspace{-0.3cm}
\end{figure}

Fig.~\ref{PhysicalLayoutPromptFig} shows the example of extracted physical layout prompt of OA333X1 layout. 
From the Fig.~\ref{PhysicalLayoutPromptFig}(a), the unit of x coordinate is the half contacted-poly-pitch (CPP), and the unit of y coordinate is half cell row. As a result, there are 29 columns and 2 rows in OA333x1 layout in the extracted coordinate based physical layout prompt in Fig.~\ref{PhysicalLayoutPromptFig}(b).
For each coordinate, we show the net name, placed device, and the terminals (i.e., source, drain, gate) of the placed device. 
The net name, and device are dummy when there are no devices in the netlist being placed at the coordinate.
Here, the column-based physical layout report format helps LLM identify the common gate and diffusion connections of PMOS and NMOS.

\subsection{Routability report prompt}
\begin{figure} [!h]
	\centering
	\includegraphics[width=0.8\columnwidth]{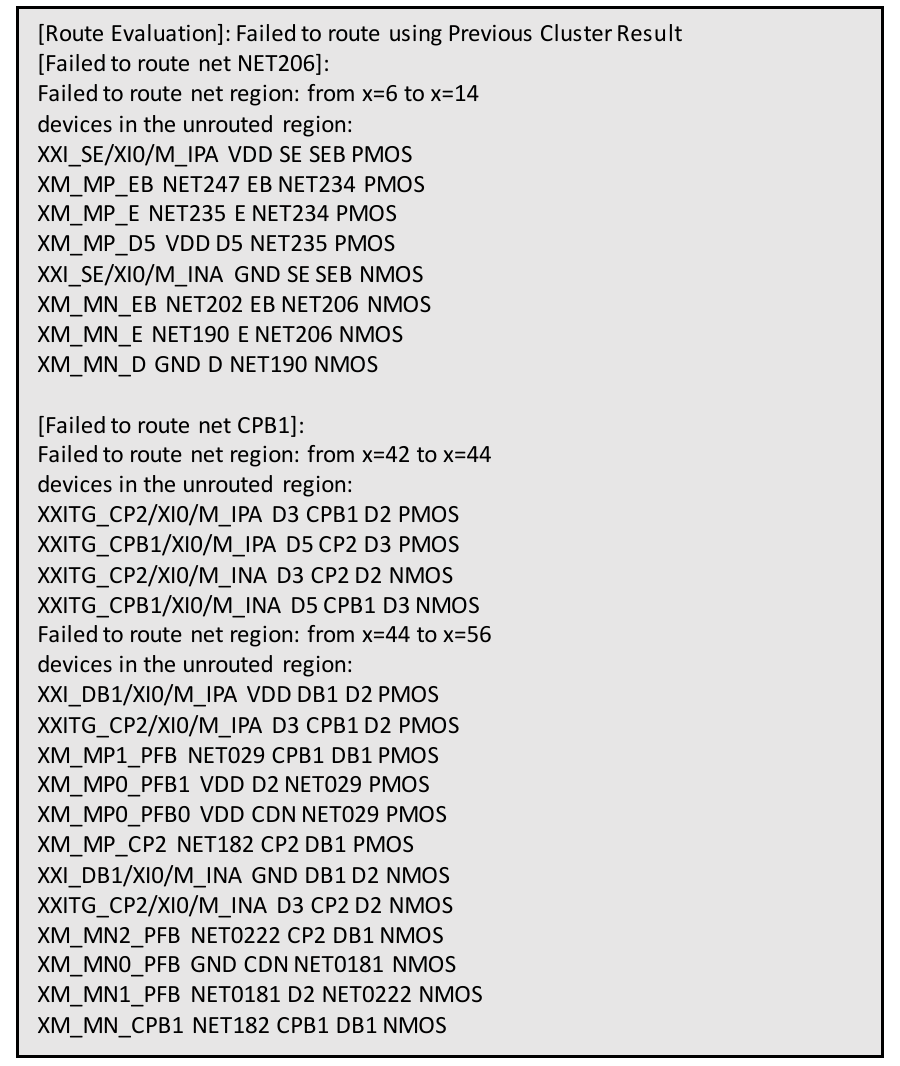}
	\caption{An example of extracted routability report prompt of Seq13 in Table~\ref{CellWidthTable}.}
	\label{RoutabilityPromptFig}
    \vspace{-0.3cm}
\end{figure}

We select the Seq13 in Table~\ref{CellWidthTable} for describing the routability report prompt format since there are no unrouted nets in OA333X1 layout as shown in Fig.~\ref{RoutabilityPromptFig}.
In the routability report, we provide information about the unrouted nets, the corresponding pairs of x-coordinates of net terminals, and the placed devices inside the unrouted region. These placed devices within the unrouted region offer insight into routing congestion and required transistor pin access, assisting LLM in identifying potential good cluster constraints to improve routability. For example, if routing congestion or pin density is too high in an unrouted net region, leveraging common transistor terminal sharing across PMOS and NMOS, as well as diffusion sharing, can reduce pin density and routing resource usage by creating cluster constraints that consider the high connection nets or problematic nets of transistor pins in an unrouted net region.

\end{document}